\documentclass{aa}

\def\grsim{\,\lower 1mm \hbox{\ueber{\sim}{>}}\,}
\def\lesssim{\,\lower 1mm \hbox{\ueber{\sim}{<}}\,}
\def\ueber#1#2{{\setbox0=\hbox{$#1$}%
  \setbox1=\hbox to\wd0{\hss$ #2$\hss}%
  \offinterlineskip
  \vbox{\box1\box0}}{}}
\usepackage{graphicx}

\begin{document}

%
   \title{Distribution of dark and baryonic matter in clusters of galaxies}


   \author{ A. Castillo-Morales\inst{1,3},
           S. Schindler\inst{2,3}}
           
   \offprints{A. Castillo-Morales, acm@ugr.es}

   \institute{Dpto. F\'{\i}sica Te\'orica y del Cosmos,       
              Universidad de Granada,                        
              Avda. Fuentenueva s/n,                         
              18071 Granada,                                 
              Spain             
\and
     Institut f\"ur Astrophysik,
     Universit\"at Innsbruck,  
     Technikerstr. 25, 
     A-6020 Innsbruck, Austria                                   
\and
Astrophysics Research Institute,              
               Liverpool John Moores University,             
	       Twelve Quays House,                           
               Birkenhead CH41 1LD,                          
               United Kingdom}

   \date{\today}

\abstract{
We present the analysis of baryonic and non-baryonic matter distribution in a sample of ten nearby clusters ($0.03<z<0.09$) with temperatures between 4.7 and 9.4 keV. These galaxy clusters are studied in detail using X-ray data and global physical properties are determined. Correlations between these quantities are analysed and compared with the results for distant clusters. We find an interesting correlation between the extent of the intra-cluster gas relative to the dark matter distribution. The extent of the gas relative to the extent of the dark matter tends to be larger in less massive clusters. This correlation might give us some hints on non-gravitational processes in clusters. We do not see evolution in the gas mass fraction out to a redshift of unity. Within $r_{500}$, the mean gas mass fraction obtained is $(0.16\pm0.02)h_{50}^{-3/2}$. 

}
\titlerunning {Distribution of dark and baryonic matter}
   \maketitle

%

\section{Introduction}

Clusters of galaxies can be regarded in many respects as representative for the
universe as a whole. As clusters accumulate mass from a large volume the
baryon fraction in clusters is representative for the baryon fraction
in the universe and hence can be used to determine
$\Omega_{\rm m}$ when comparing the cluster baryon fraction to the upper
limit of $\Omega_{\rm baryon}$ from primordial nucleosynthesis.

The total mass of a cluster can be determined in various independent
ways: strong 
and weak lensing, X-ray observations, galaxy velocities and the
Sunyaev-Zel'dovich effect. From these mass
determinations an average baryon fraction of 15-20\% was inferred 
(e.g Mohr et al.~\cite{Mohr99}; Ettori \& Fabian~\cite{Ett99}; Arnaud \& Evrard~\cite{Arn99}; Grego et al.~\cite{Grego})
which implies an upper limit of the mean matter density 
$\Omega_{\rm m} \lesssim0.3 - 0.4$.

Furthermore, the distribution of the different cluster components can
be studied when mass profiles are available. From X-ray
observations the gas density and the gas temperature can be
determined, which yield not only the gas mass profile, but with the
assumptions of 
hydrostatic equilibrium and spherical symmetry also the total mass profile of the cluster. 
Numerical simulations showed that these masses are quite
reliable for virialised clusters (Evrard et al.~\cite{Ev96}; Schindler~\cite{Sch96}; Dolag \& Schindler~\cite{Do00}). A
comparison of both radial profiles gives information on the relative distribution of dark and baryonic matter. This can be used to learn about the cluster formation process: whether all the energy comes only from
the gravitational collapse, in which case both distributions baryonic
and dark matter are expected to have the same distributions, or
whether there are additional physical processes involved.

Deviations of the $L_{\rm X}- T$ relation compared to the relation expected for self-similar
scaling (e.g. Arnaud \& Evrard~\cite{Arn99}) and
entropy studies (Ponman et al.~\cite{Pon99}) suggested that at least for small
galaxy systems additional (pre-)heating processes (Tozzi~\cite{Toz01})
play an important role.
Numerical simulations were performed to test this suggestion (Bialek
et al.~\cite{Bial01}, Borgani et al.~\cite{Borg01})
finding that an entropy floor around 50 - 100 keV/cm$^2$ is required
to fit the observational results.

In this article we determine various cluster properties for massive systems and compare the distribution of the different components within these clusters.
Moreover, we use the determined
masses to derive relations between the masses and other cluster
quantities, which give more information about cluster formation and
evolution.

We use a sample of nearby clusters with the best ROSAT and ASCA
data. This sample is complemented with four new observations of distant clusters performed by Chandra and XMM.

Throughout this paper we use $\rm{H}_0 = 50$ km/s/Mpc and $\rm{q}_0=0.5$.


\section{The sample}

Since the aim of this paper is the analysis of the total and gas
mass distribution in nearby galaxy clusters we selected clusters in which an
accurate total mass determination is possible, i.e with relaxed and
symmetric morphologies, good temperature measurements and well determined
surface brightness profiles. Therefore we obtain a high quality sample that consists of ten best clusters in the redshift range $0.03< z <0.09$ observed with the ROSAT PSPC. Obviously, the sample 
is not complete in any sense, therefore no analyses of distribution
functions can be made with it. But the analyses of correlations
between the different quantities are not affected by the
incompleteness (see Finoguenov et al.~\cite{Fin01}). Cluster
temperatures measured by the ASCA satellite are taken from Markevitch et al.~(\cite{Mark98}), Markevitch et al.~(\cite{Mark99}), Sarazin et al.~(\cite{Sar98}) and Bauer \& Sarazin~(\cite{Bau00}) resulting in a temperature range of 4.7-9.4 keV.
In Table~\ref{tbl:properties} the different properties for each cluster are listed.

Of all the new observations only the best data are selected in order to obtain the most reliable X-ray mass estimate. In Table~\ref{tbl:distant} the published data we use to derive masses are listed.


\section{Data analysis}
	
We use X-ray imaging data retrieved from the ROSAT archive\footnote{\footnotesize{http://www.xray.mpe.mpg.de/rosat/archive}} to
determine the surface brightness profiles of the clusters. For each
cluster a ROSAT PSPC image was reduced using the standard analysis
with the EXSAS software. In order to maximize the signal-to-noise ratio,
we choose the hard energy band (0.5-2.0 keV) corresponding to channel
numbers 52-201.

The images were corrected for exposure variations and telescope
vignetting using exposure maps.
The cluster cores are somewhat blurred in PSPC images by the point spread function (PSF)
($\approx 20-30$ arcsec FWHM for ROSAT/PSPC). This effect is more important for compact clusters. For example for the compact cluster A780, Neumann $\&$ Arnaud~(\cite{Neu99}) estimated the core radius to be overestimated by 
about 10$\%$. For the other clusters in our sample, which have a
larger core radius, we estimate that the effects of the PSF are much
smaller than the statistical errors. Therefore no correction for the PSF is necessary.

We generate radial surface brightness profiles in 
concentric annuli (centred on the emission maximum of the
cluster) excluding obvious point sources manually. The observed
profiles are fitted with a $\beta$-model, (Cavaliere $\&$ Fusco-Fermiano~\cite{Cav76}) plus background.

\begin{equation}
S(b)=S_{0}\left[1+\frac {b^{2}}{r_{\rm c}^{2}}\right]^{-3\beta+\frac {1}{2}}+B.
\label{eq:surf}
\end{equation}

The parameters $r_{\rm c}$ (core radius), $\beta$, $S_{0}$ (central surface brightness) and the
background B are obtained from a least-squares fit to the X-ray
profile. The slope $\beta$ and the core radius $r_{\rm c}$ are not
independent parameters, with $\beta$ increasing when
$r_{\rm c}$ increases. They are found to
range between 0.6 and 0.8 and between 130 and 290 kpc, respectively.

However, the overall $\beta$-model fit is a poor description of the central
region of some clusters where excess emission is observed. Indeed we find in most of the 
cases very large $\chi^{2}$ values when fitting the entire
cluster emission. We reduce the $\chi^{2}$ values by excluding the central bins from the fit. 
The best 
fit $\beta$ -model was determined excluding the data within the
cooling radius taken from Peres et al.~(\cite{Per98}), Allen $\&$ Fabian~(\cite{All97}), White et al.~(\cite{Wh97}). 
For the clusters in our sample with excess central emission, the exclusion of
the central part of the profile yields larger $\beta $ values and
core radii values compared to the overall $\beta$-model fit. Excluding the central excess we underestimate the gas density at the centre (about 12\% for the central 3$'$ in the case of the cluster A2199). However we estimate that the central gas mass contributes only with a few percent in the gas mass at larger radius (about 3\% for cluster A2199 at the radius of $r_{500}$). Therefore for the gas mass determination, this underestimate at small radii is negligible when we integrate the gas density out to large radii.

In Table~\ref{tbl:parameters} the resulting parameters are shown. The reported errors are 90$\%$ confidence level.

\begin{table*}[t]
\caption {In Cols. 1 and 2 the clusters and their redshift are shown. In Col. 3 the hydrogen column density from Dickey \& Lockman~(\cite{Dick90}) is listed. The temperatures in Col. 4 are ASCA emission-weighted temperatures excluding cooling flows
and other contaminating components. Col. 5 lists the ratio of the bolometric flux to the ROSAT PSPC countrate in the energy range 0.5-2.0 keV using the programme PIMMS (http://www.heasarc.gsfc.nasa.gov/Tools/w3pimms.html). ASCA temperatures are taken from references: 1$-$Markevitch et al. (1998); 2$-$Markevitch et al. (1999); 3$-$Sarazin et al. (1998); 4$-$Bauer \& Sarazin (2000).}

\begin{center}
\begin{tabular}{ c  c  c  c  c  c }
\hline
Cluster&$z$&$N_{\rm H}$&$kT_{\rm gas}$&$F_{\rm Xbol}/ \rm countrate$&Ref.\\ 
    &  &\footnotesize($10^{20} \rm{cm^{-2}}$)&\footnotesize(keV) &\footnotesize($10^{-11}\rm{erg/counts/cm^{2}}$)\\
\hline\hline
\\
A496 & 0.033& 4.58& 4.7$\pm$0.2             &4.3& 2\\               
A2199& 0.030& 0.863 & 4.8$\pm$0.2            &3.9& 2\\  
A3112& 0.075& 2.61& $5.3^{+0.7}_{-1.0}$     &4.4& 1\\ 
A1651& 0.085& 1.81& 6.1$\pm$0.4             &4.5& 1\\
A3571& 0.040& 3.71 & 6.9$\pm$0.2            &4.9& 1\\  
A1795& 0.062& 1.19& 7.8$\pm$1.0             &4.9& 1\\  
A401 & 0.075& 10.5& 8.0$\pm$0.4            &6.0& 1\\  
A478 & 0.088& 15.1& $8.4^{+0.8}_{-1.4}$    &6.8& 1\\   
A644 & 0.070& 6.82& $8.6^{+0.7}_{-0.6}$     &5.8& 4\\   
A2029& 0.077& 3.04& $9.4^{+0.6}_{-0.5}$     &5.6& 3\\

\\ \hline
\end{tabular}
\label{tbl:properties}
\end{center}
\end{table*}

\begin{table*}[t]
\caption {Published data from XMM and Chandra observations used to derived the total and gas mass. In Cols. 1 and 2 the clusters and their redshift are listed. The gas temperature is shown in Col. 3. The published fit parameters of the  $\beta$ model are listed in Cols. 4, 5 and 6: the slope $\beta$, the core radius $r_{\rm c}$ and the central electron density $n_{\rm e0}$. The quoted errors are 90$\%$ confidence level for all quantities except for the parameters in cluster RX-J0849+4452 and the value of the temperature in A1835 which are 68$\%$ confidence level. References: (1) Standford et al.~(\cite{Stan01}); (2) Schindler et al.~(\cite{Sch01}); (3) Arnaud et al.~(\cite{Arn02}); (4) Majerovitch et al.~(\cite{Maj02}).}
\begin{center}
\begin{tabular}{ c  c  c  c  c c  c }
\hline
 Cluster& z &$kT_{\rm gas}$& $\beta$&$r_{\rm c}$&$n_{\rm e0}$&Ref. 
\\
 & & {\footnotesize(keV)}& &{\footnotesize(kpc)}& ($10^{-2}\rm{cm^{-3}}$) & \\ 
\hline\hline
\\
RX-J0849+4452 & 1.26 & $5.8^{+2.8}_{-1.7}$ &$0.61\pm0.12$ & $100.2\pm30.7$ &1.42 &Chandra (1) \\
RBS797        & 0.354& $7.7^{+1.2}_{-1.0}$ &$0.63\pm0.01$ & $49\pm4$   &8.86 &Chandra (2) \\
RX-J1120.1+4318&0.6  & $5.3\pm0.5$         &$0.78^{+0.06}_{-0.04}$ & $201^{+27}_{-18}$       &0.81 & XMM (3) \\
A1835         & 0.25 & $7.6\pm0.4$         &$0.704\pm0.005$& $202.3\pm7.1$ &1.47 &XMM (4) \\
\\ 
\hline
\end{tabular} 
\end{center}
\label{tbl:distant}
\end{table*}

\begin{table*}[t]
\caption {X-ray quantities as measured from ROSAT/PSPC and ASCA observations. The clusters are listed in Col. 1. Col. 2 shows the emission-weighted gas temperature. In Cols. 3, 4, 5 and 6 the fit parameters of the $\beta$ model are
shown: the slope $\beta$, the core radius $r_{\rm c}$, the central surface
brightness $S_{0}$ and the background 
$B$ in the energy band 0.5 - 2.0 keV. $S_{0}$ is in units of $10^{-2}$
ROSAT/PSPC $\rm{counts/s/arcmin^{2}}$. B is in units of $10^{-4}$
ROSAT/PSPC $\rm{counts/s/arcmin^{2}}$. In Col. 7 $r_{\rm f}$ denotes the radius
range fitted.}
\begin{center}
\begin{tabular}{ c  c  c  c  c c c }
\hline
 Cluster&$kT_{\rm gas}$&$\beta$&$r_{\rm c}$&$S_{0}$&B&$r_{\rm f}$ 
\\
 &{\footnotesize(keV)}& &{\footnotesize(kpc)}& & &{\footnotesize (Mpc)}\\ 
\hline\hline
\\
A496 & 4.7$\pm$0.2             &$0.64\pm0.02$ & $185\pm16$ & 4.0  & 4.4& 0.14 - 2.9\\
A2199& 4.8$\pm$0.2             &$0.64\pm0.01$ & $132\pm7$  & 9.4  & 2.4& 0.15 - 2.8\\
A3112& $5.3^{+0.7}_{-1.0}$     &$0.65\pm0.03$ & $181\pm46$ & 6.6  & 2.9& 0.27 - 5.8\\
A1651& 6.1$\pm$0.4             &$0.68\pm0.03$ & $224\pm24$ & 7.3  & 3.5& 0.13 - 6.4\\
A3571& 6.9$\pm$0.2             &$0.66\pm0.02$ & $235\pm20$ & 7.1  & 4.4& 0.14 - 2.8\\
A1795& 7.8$\pm$1.0             &$0.68\pm0.01$ & $197\pm16$ & 11.3 & 2.8& 0.19 - 2.8 \\
A401 & 8.0$\pm$0.4             &$0.67\pm0.02$ & $284\pm21$ & 5.4  & 2.5& 0.11 - 5.7\\
A478 & $8.4^{+0.8}_{-1.4}$     &$0.69\pm0.02$ & $201\pm20$ & 11.7 & 1.5& 0.22 - 6.1 \\
A644 & $8.6^{+0.7}_{-0.6}$     &$0.72\pm0.02$ & $239\pm17$ & 6.9  & 2.3& 0.20 - 5.4\\
A2029& $9.4^{+0.6}_{-0.5}$     &$0.68\pm0.02$ & $244\pm24$ & 9.7  & 5.1& 0.29 - 5.9\\
\\ 
\hline
\end{tabular} 
\end{center}
\label{tbl:parameters}
\end{table*}

\section{Mass determination}

Once we deproject the surface brightness to three-dimensional 
density with the $\beta$ model, together with the
assumptions of hydrostatic equilibrium and spherical symmetry the integrated mass within radius
$r$ can be determined as

\begin{equation}
M(r) = {-kr\over \mu m_p G} T(r) \left({ d \ln \rho(r) \over d \ln r }+
                                    { d \ln T(r)    \over d \ln r }\right),
\label{eq:mass}
\end{equation}
with $\rho$ and $T$ being the density and the temperature of the
intra-cluster gas, respectively. 
$k$, $\mu$, $m_p$, and $G$ are the
Boltzmann constant, the molecular weight, the proton mass, and
the gravitational constant.

\begin{table*}[t]
\caption {Results of the isothermal analysis: total mass, gas mass and gas mass
fraction for the cluster sample. The first column gives the cluster name. Col. 2 denotes the radius $r_{500}$ which 
comprises an overdensity of 500 over the critical density. Cols.
3, 4 and 5 list the total mass, the gas mass and the gas mass
fraction within $r_{500}$, respectively. In Cols. 6, 7 and 8
the same quantities are listed for a radius $0.5\times r_{500}$. For all the quantities the reported errors are 90$\%$ confidence level.}                                                                     
\begin{center}
\begin{tabular}{ l  l  l  l l l l l}
\hline
 Cluster& $r_{500}$&$M_{\rm tot}(r_{500})$&$M_{\rm gas}(r_{500})$&$f_{\rm gas}(r_{500})$& $M_{\rm tot}(\frac{r_{500}}{2})$&
$M_{\rm gas}(\frac{r_{500}}{2})$&$f_{\rm gas}(\frac{r_{500}}{2})$\\ 
  &{\small(Mpc)}&{\small($10^{14}M_{\odot}$)}&{\small($10^{14}M_{\odot}$)}& &{\small($10^{14}M_{\odot}$)}&{\small($10^{14}M_{\odot}$)}& \\
\hline\hline
\\
A496 &$1.42\pm0.03$         &$4.6\pm0.5$ &0.69       &$0.150\pm0.019$  &$2.2\pm0.3$ &0.27 &$0.12\pm0.03$\\
A2199&$1.45\pm0.03$         &$4.8\pm0.6$ &0.68       &$0.142\pm0.018$  &$2.4\pm0.3$ &0.28 &$0.12\pm0.03$\\ 
A3112&$1.44^{+0.09}_{-0.14}$&$5.3^{+0.9}_{-1.2}$&0.94&$0.18^{+0.04}_{-0.03}$&$2.5^{+0.4}_{-0.5}$&0.37&$0.15^{+0.07}_{-0.05}$\\
A1651&$1.55\pm0.05         $&$6.9\pm0.9$ &1.24       &$0.18\pm0.02$  &$3.2\pm0.4$   &0.50 &$0.16\pm0.04$\\
A3571&$1.73^{+0.03}_{-0.02}$&$8.5\pm1.0$ &1.50       &$0.18\pm0.02$  &$4.0\pm0.5$ &0.60 &$0.15\pm0.04$\\
A1795&$1.82^{+0.12}_{-0.13}$&$10.4\pm1.8$&1.49       &$0.14\pm0.02$  &$5.0\pm0.9$   &0.64 &$0.13\pm0.05$\\
A401 &$1.78^{+0.04}_{-0.05}$&$10.2\pm1.2$&1.93       &$0.19\pm0.02$  &$4.7\pm0.6$   &0.75 &$0.16\pm0.04$\\
A478 &$1.83^{+0.09}_{-0.16}$&$11.5^{+1.7}_{-2.3}$&1.83&$0.16^{+0.03}_{-0.02}$&$5.5^{+0.8}_{-1.1}$&0.80&$0.14^{+0.06}_{-0.04}$\\
A644 &$1.94\pm0.07         $&$13.0^{+1.8}_{-1.7}$&1.52&$0.117\pm0.017$&$6.2^{+0.9}_{-0.8}$&0.68&$0.11\pm0.03$\\
A2029&$1.95^{+0.06}_{-0.05}$&$13.4\pm1.7$&2.16&$0.16\pm0.02$&$6.4\pm0.8$&0.90&$0.14\pm0.04$\\
\\
\hline
\end{tabular} 
\end{center}
\label{tbl:masses}
\end{table*}

\subsection{Isothermal analysis}

As well as the density profiles we need temperature profiles to
determine the total cluster mass. We follow two different approaches for the temperature profile. In our first study we assume isothermality. 
In Sect.~\ref{section:tempgrad} we include the temperature gradients derived from 
Markevitch's ASCA analysis to see how the total cluster mass is affected.
In the isothermal approach we neglect the term associated with the temperature gradient in Eq.~(\ref{eq:mass}). In this case the total mass profile is:
\begin{equation}
M(<r)=\frac{3kT_{\rm gas}r^{3}\beta}{G\mu m_{\rm p}}(\frac{1}{r^{2}+r_{\rm c}^{2}}),
\label{eq:M}
\end{equation}
where the mean atomic weight $\mu$ is assumed to be 0.61. 

The total mass thus depends linearly on both $\beta$ and
$T_{\rm gas}$. In the case of clusters with central X-ray excess, we use emission-weighted gas temperature obtained by excluding the central part of the cluster. Typical mass profiles are shown in Fig.~\ref{fig:mass} for the cluster Abell 2199.

After having determined the gravitational mass profiles for the clusters, it is important to fix the radius within which the cluster masses can be calculated. As the mass of a cluster increases
with radius, masses can only be compared when derived within
equivalent volumes. Simulations by Evrard et al.~(\cite{Ev96}) showed that the assumption of hydrostatic equilibrium is generally valid 
within at least radius $r_{500}$, where the mean gravitational mass density
is equal to 500 times the critical density
$\rho_{\rm c}(z)=3H_{0}^{2}(1+z)^{3}/8 \pi G$. The resulting cluster masses and gas masses within
$r_{500}$ and $0.5\times r_{500}$ for all the clusters are listed in
Table~\ref{tbl:masses}.

The errors on the total mass can be estimated as follows.
The total cluster mass is affected by the errors in the parameters of the
$\beta$-model. This error is estimated to be about 5\%. The uncertainties in the total mass estimate are much larger due to the uncertainty in temperature estimates and possible temperature gradients. For larger radii we assume an error of 10\% in the total cluster mass, caused by the existence of a temperature gradient.
There are also additional uncertainties coming from deviations from spherical symmetry (Piffaretti et al.~\cite{Piff}), deviations from hydrostatic equilibrium and projection effects (together about $15\% - 30\%$, Evrard et al.~\cite{Ev96}, Schindler~\cite{Sch96}), but these are hard to quantify for each cluster
individually. As only well relaxed clusters where chosen, these errors should be relatively small ($<15\%$) in the clusters of this sample.

We compute the error in the total cluster mass as the convolution of the error coming from the uncertainty in the fit parameters, the error in the temperature and the 10$\%$ error coming from the assumption of non isothermality. However we keep in mind that the true uncertainty of the total mass is probably greater than this value.

In the energy range considered, the gas mass estimate is almost independent on the temperature measurement. We estimate the error in the gas mass to about 5$\%$ due to the uncertainty in the fit parameters.

\subsection{Temperature gradients}
\label{section:tempgrad}

\begin{figure}[ht]
\centering
\includegraphics[width=10cm]{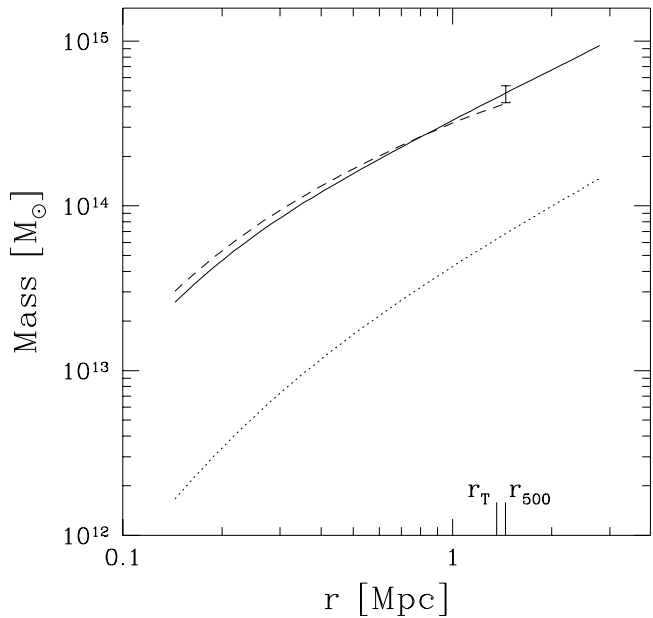}
\caption {Total mass profile assuming isothermality (solid line) and gas mass profile
(dotted line) for the cluster A2199. The dashed line represents the total mass 
profile derived with a temperature gradient. $r_{500}$ is
the radius which comprises an overdensity of 500 over the critical
density. $r_{\rm T}$ represents the maximum radius out to which the temperature profile is calculated. A typical error bar for the isothermal total cluster mass is shown at $r_{500}$. }
\label{fig:mass}
\end{figure}

\begin{figure}[ht]
\centering
\includegraphics[width=10cm]{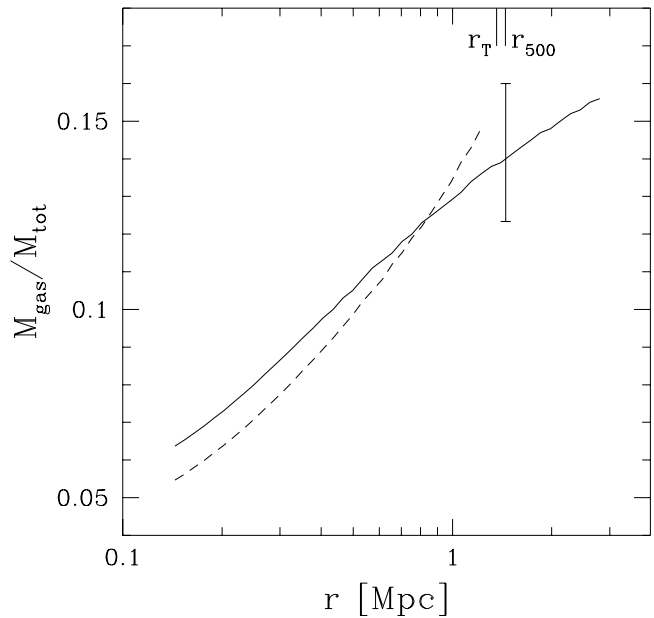}
\caption {Gas mass fraction assuming isothermality (solid line) for cluster A2199. Dashed line represents the gas mass fraction derived with a gradient of temperature. A typical error bar for the isothermal gas mass fraction is shown at $r_{500}$.}
\label{fig:fgas}
\end{figure}

The isothermal assumption may give poor estimates of the total mass if
strong temperature gradients are present. Observations with ASCA
suggest that the temperature does decrease with radius.
To estimate the effect of these temperature gradients on cluster masses, we have
calculated $M_{\rm tot}$ using the temperature profiles derived by Markevitch et al.~\cite{Mark98}, Markevitch et al.~\cite{Mark99}, Sarazin et al.~\cite{Sar98} and Bauer \& Sarazin~\cite{Bau00}. In some clusters, the temperature outside the
cooling core can be well approximated by a polytropic function

\begin{equation}
T_{\rm gas}\propto\rho_{\rm gas}^{\gamma-1}
\label{eq:polyT}
\end{equation}
where $\gamma$ is the polytropic index. In
this case the total mass enclosed in a sphere of radius $r$ (Eq.~(\ref{eq:mass})) is:
\begin{equation}
\label{eq:gradmass}
M(<r)=\frac{3\beta\gamma r^{3}}{G
m_{\rm p}\mu}\frac{kT_{\rm gas}(r)}{(r^{2}+r_{\rm c}^{2})}.
\end{equation}
Here $T_{\rm gas}$ is the true temperature, rather than a projection on to the
plane of the sky. Markevitch et al.~(\cite{Mark99}) showed that as long as
the temperature is proportional to a power of the density, and the density 
follows a $\beta$-model, the true temperature differs from the
projected temperature only by a constant factor, given by:
\begin{equation}
\label{eq:factor}
\frac{T_{\rm proj}}{T_{\rm true}}=\frac{\Gamma\left[\frac{3}{2}\beta(1+\gamma)-\frac{1}{2}\right]\Gamma(3\beta)}{\Gamma\left[\frac{3}{2}\beta(1+\gamma)\right]\Gamma(3\beta-\frac{1}{2})}.
\end{equation}

In our analysis, the projected temperature profiles are parametrized using a
linear function of the form
\begin{equation}
T_{\rm gas}(r)=T(0)-\alpha r.
\label{eq:linearT}
\end{equation}
The parameters T(0) and $\alpha$ are determined by fitting a straight
line to the projected temperature profiles  mentioned before, excluding the central cluster region in the case of clusters with central X-ray excess. It is not useful to fit a more complex function because the temperatures for consecutive annuli are determined with low accuracy.

We do not deprojected the temperatures for the mass analysis with
temperature gradient. Although this introduces some additional
uncertainty in the total mass we show below that the results using
true temperatures or projected temperatures are very similar.

For the cluster A2199 we compare the total mass
derived using the linear gradient Eq.~(\ref{eq:linearT}) and the
mass calculated with the polytrope Eq.~(\ref{eq:polyT}) by Markevitch et al.~(\cite{Mark99}). The estimated values for the total cluster mass agree within the errors. There is a difference $\leq 10\%$ at large radii ($r>1Mpc$) and
$\leq 15 \%$ at the small radius ($r=0.2 Mpc$). For example Markevitch et al.~(\cite{Mark99}) using a polytrope 
equation with $\gamma=1.17$ find a total cluster mass of
$(0.65\pm0.11)\times10^{14}M_{\odot}$, $(2.9\pm0.3)\times10^{14}M_{\odot}$ and
$(3.6\pm0.5)\times10^{14}M_{\odot}$ at a radius of 0.2 Mpc, 1 Mpc and
$r_{500}=1.3$ Mpc respectively, for cluster A2199. We derive the total cluster mass for
this cluster using the linear gradient $T(r)=-2.1r+5.6$ where $\it{r}$ is
units of Mpc and temperature is in keV. The masses obtained in this
case are quite similar, $(0.56\pm0.14)\times10^{14}M_{\odot}$, $(3.1\pm0.8)\times10^{14}M_{\odot}$ and $(3.9\pm1.0)\times10^{14}M_{\odot}$ at a radius of 0.2 Mpc, 1 Mpc and 1.3 Mpc, respectively.  
Therefore we conclude that for our purpose of comparison, it is sufficient to use a linear gradient of projected temperatures.

\begin{figure}[ht]
\centering
\includegraphics[width=10cm]{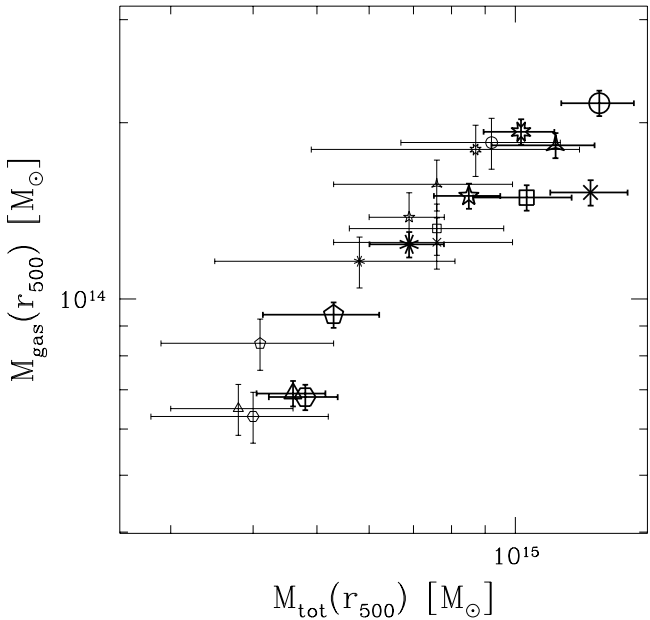}
\caption {Comparison of masses obtained with the
isothermal and temperature profile analyses at radius $r_{500}$. The
smaller symbols represent the temperature profile analysis.}
\label{fig:compareM}
\end{figure}

\begin{figure}[ht]
\includegraphics[width=10cm]{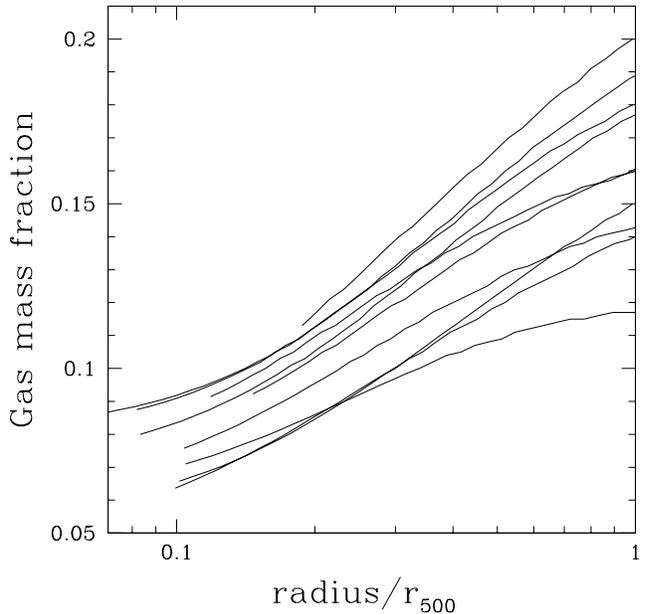}
\caption[] {Gas mass fraction profiles derived for the nearby cluster sample. Profiles are plotted from the minimum radius fitted in the $\beta$ model.}
\label{fig:fgasr500}
\end{figure}

In the following we will apply only a linear temperature gradient. In this case we estimate the errors in the total cluster mass and gas mass as follows. Due to the errors in the temperature, the errors in the linear fit are large. We estimate the errors in the total cluster mass using the different possible temperature gradients given by the lower and upper values of T(0) and $\alpha$ (see Eq.~(\ref{eq:linearT})). When the different possible temperature gradients are used, the radii $r_{r500}$ and $r_{r500}/2$ change significantly and hence the total and gas mass enclosed by them. For the gas mass an error of $\sim 10\%$ is estimated. 

In Fig.~\ref{fig:mass} the different mass
profiles calculated using the isothermal model, and a linear gradient of
temperature, for cluster A2199 can be seen. With the assumption of
constant temperature (solid line) the total mass is overestimated ($\sim 10\%$) at large radii and underestimated at small radii,
compared to the temperature gradient analysis (dashed line). The overestimate of the total mass is only significant at radii larger than $r_{\rm T}$, with $r_{\rm T}$ being the radius out to which the temperature was measured. This trend is
observed for all the clusters in our sample. Fig.~\ref{fig:mass} also shows the gas mass profile (dotted line) in the radius range that has
been used to fit the data with the $\beta$-model.

Fig.~\ref{fig:fgas} shows the gas mass fraction, defined as the ratio between the gas mass $M_{\rm gas}$ and the total mass $M_{\rm tot}$. The difference in the gas mass fraction between the
isothermal and non-isothermal cluster studies is shown as well (solid and dashed lines, respectively). At small
radii both analyses provide approximately parallel profiles with
smaller values obtained in the temperature gradient case. At larger radii the temperature profile
analysis yields steeper profiles and  higher values ($\sim 10\%$) compared to the isothermal one. The gas mass fraction derived using a gradient of temperature is not reliable
beyond the radius $r_{\rm T}$ where the gradient is determined and thus it 
is not plotted.

\begin{figure}[h]
\centering
\includegraphics[width=7cm]{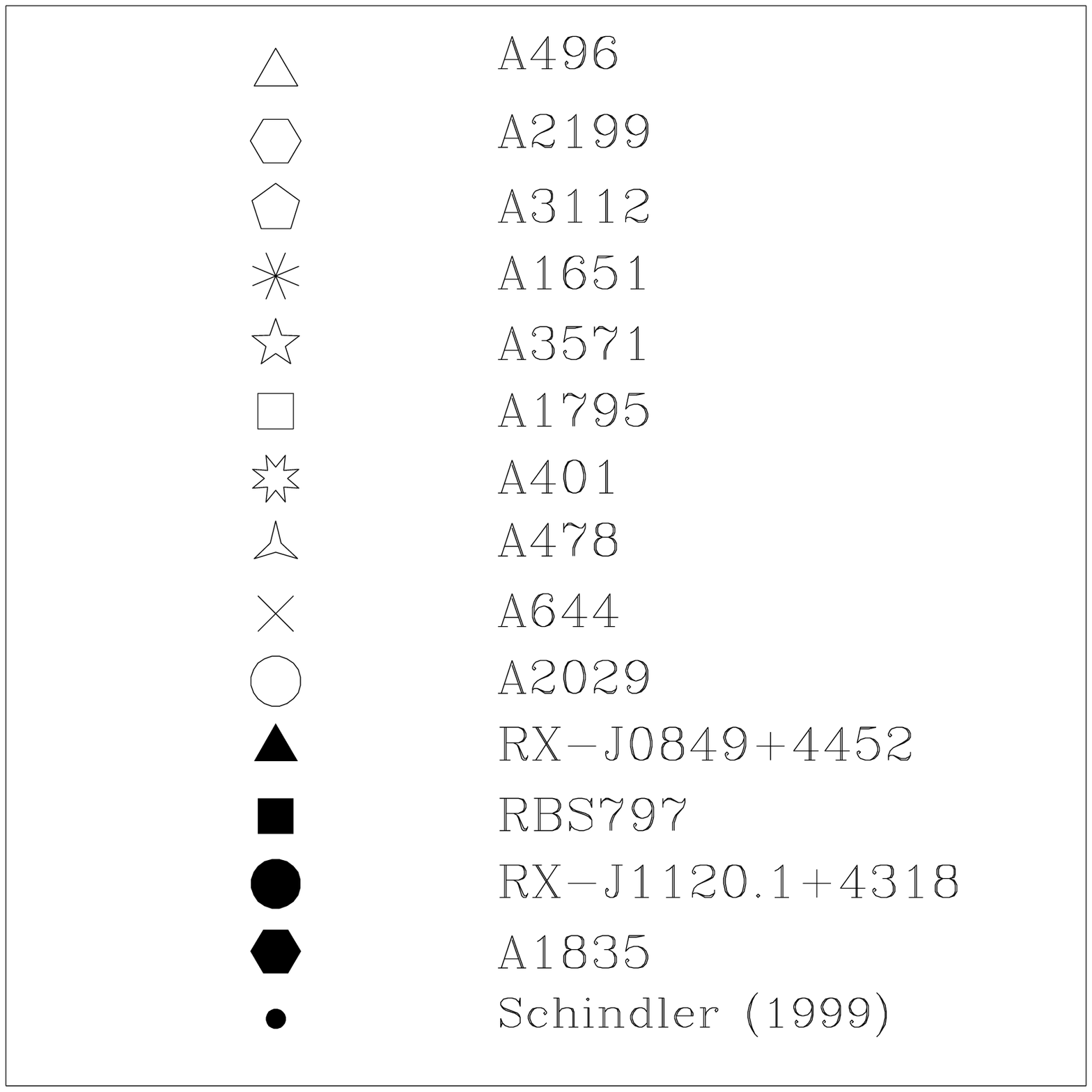}
\caption {Symbols used for the various clusters in
the figures.}
\label{fig:symbols}
\end{figure}

The masses calculated with both analyses
(isothermal and temperature profile) are compared in Fig.~\ref{fig:compareM} for each cluster in the
sample at radius $r_{500}$. The symbols correspond to the cluster name listed in
Fig.~\ref{fig:symbols}. The smaller symbols represent the values for
the temperature profile analysis. Although the gas mass profile is not
influenced by the change in the cluster temperature, the gas mass at
radius $r_{500}$ is different in the two analyses due to the difference in
$r_{500}$ which depends on the temperature.

\begin{figure}[ht]
\centering
\includegraphics[width=10cm]{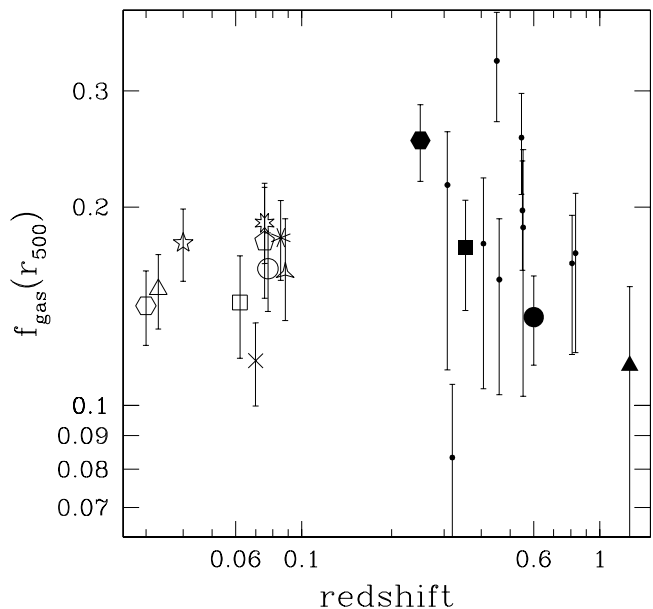}
\caption {Gas mass fraction versus redshift. No clear evolution of the gas mass fraction can be seen out to a redshift of unity. See Fig.~\ref{fig:symbols} for the different cluster symbols.}
\label{fig:fgas_z}
\end{figure}

\begin{figure}[ht]
\centering
\includegraphics[width=10cm]{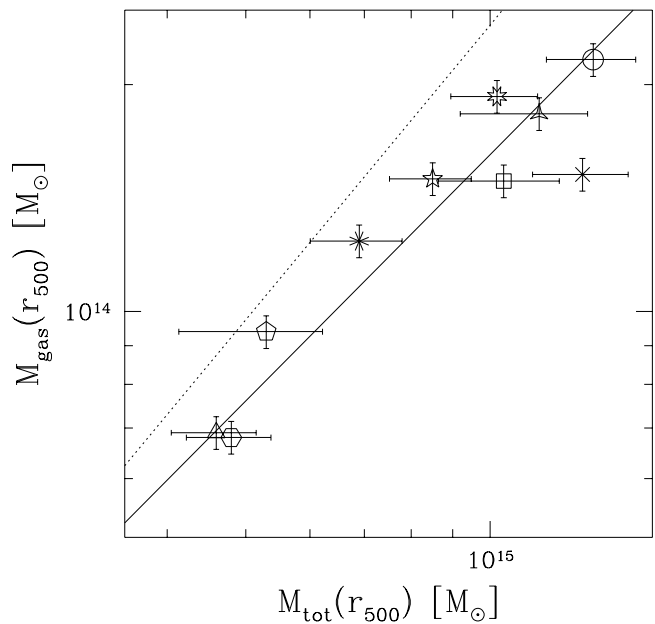}
\caption {Gas mass versus total mass. In solid line is shown the best fit for the nearby sample. For comparison, in dotted line is plotted the best fit obtained for the distant sample by Schindler~(\cite{Sch99})}
\label{fig:relMgMt}
\end{figure}

\section{Results and discussion}
In the following all the results presented are refered to the analysis of the cluster sample assuming isothermality.
\subsection{Gas mass fraction}

The gas mass fraction defined as
$f_{\rm gas}=M_{\rm gas}/M_{\rm tot}$ is calculated for each cluster in
the sample. The errors associated with the gas mass fractions are calculated by
the convolution of errors in the total cluster mass and the gas mass. We find that inside each cluster the gas mass fraction increases with radius (see Fig.~\ref{fig:fgasr500}) implying that the gas distribution is more extended than dark matter. 

A low $\Omega$ is required to reconcile the high gas mass fraction of $<f_{\rm gas}>_{r_{500}}=0.16\pm0.02$, whith the baryon fraction predicted by primordial nucleosynthesis. 

To test whether there is any evolution of the gas mass fraction we
plot this quantity versus redshift (Fig.~\ref{fig:fgas_z}) including the analysis from
Schindler~(\cite{Sch99}) for distant clusters (0.3$<$z$<$1.0). In this figure, as well as in all other following figures,
each cluster is plotted with a different symbol (see Fig.~\ref{fig:symbols}). We include in the comparison another distant
clusters: RX$-$J0849+4452 (z=1.26), RBS797 (z=0.354), A1835 (z=0.25) and RX$-$J1120.1+4318 (z=0.6) where we calculate
the total and gas mass using the published parameters from the Chandra and XMM
data analysis (see Table~\ref{tbl:distant}).

\begin{figure}[ht]
\centering
\includegraphics[width=10cm]{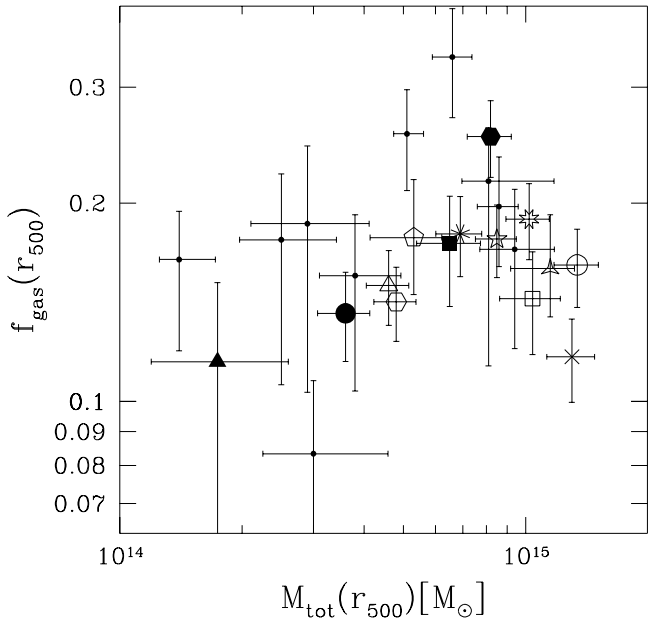}
\caption {Gas mass fraction versus total cluster mass for the nearby and distant sample. The gas mass fraction shows no clear dependence on the total mass.} 
\label{fig:relfgasMt}
\end{figure}

\begin{figure}[ht]
\centering
\includegraphics[width=10cm]{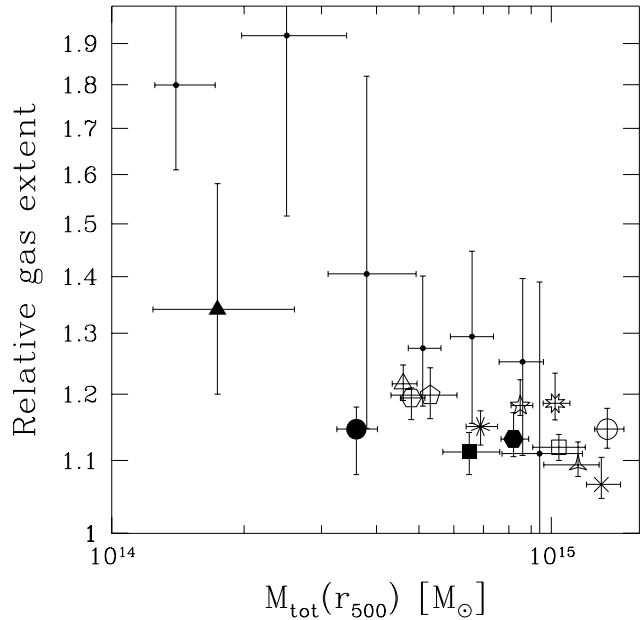}
\caption{Ratio of gas mass fraction at $r_{500}$ and $r_{500}/2$ as a
measure for the relative extent of the gas distribution versus
total cluster mass at $r_{500}$. For clarity, clusters with the largest error bars from Schindler~(\cite{Sch99}) are not shown.}
\label{fig:E_Mt}
\end{figure}

In the gas mass fraction we see no clear trend with redshift. 
The mean value in our sample is
$<f_{\rm gas}>=0.16\pm0.02$ which is in agreement with other nearby samples, Mohr et 
al.~(\cite{Mohr99}): $f_{\rm gas}$=0.21, Ettori $\&$ Fabian~(\cite{Ett99}):
$f_{\rm gas}$=0.17 and also in agreement with the value for
distant sample by Schindler~(\cite{Sch99}),
$f_{\rm gas}$=0.18. Therefore we conclude that we do not
see evolution in the gas mass fraction out to a redshift of unity. In contrast to this result Ettori $\&$ Fabian~(\cite{Ett99}) found indications for a decrease of $f_{\rm gas}$ with increasing redshift in a nearby sample. Matsumoto et
al.~(\cite{Matsu00}) found no clear evidence of evolution of $f_{\rm gas}$ for the
clusters at z $<$ 1.0.

In Fig.~\ref{fig:relMgMt} we compare the gas mass with the total
cluster mass. A trend of an increasing gas mass with total mass is visible in our
sample. This trend was also found for distant clusters by Schindler~(\cite{Sch99}) (dotted line in Fig.~\ref{fig:relMgMt}). A linear fit
regression yields
\begin{equation}
M_{\rm gas}(r_{500})=0.13M_{\rm tot}(r_{500})^{(1.09 \pm 0.12)}.
\label{eq:mg-mt}
\end{equation} 
$M_{\rm gas}(r_{500})$ and $M_{\rm tot}(r_{500})$ are in units of $10^{14}M_{\odot}$. This
trend is in agreement with the analysis by Arnaud $\&$ Evrard~(\cite{Arn99}).

Because the exponent in Eq.~(\ref{eq:mg-mt}) is close to unity, we
find no clear dependence of the gas mass fraction on the total
mass in our sample (see Fig.~\ref{fig:relfgasMt} where the nearby and distant sample are shown). 

Several authors have also related measured values of $f_{\rm gas}$
and $M_{\rm tot}$ or $T_{\rm gas}$. Up to now the results are discordant concerning the form of this relation. For example, David et
al.~(\cite{Dav95}) found indications for an increase of $f_{\rm gas}$ with
increasing $T_{\rm gas}$. Allen $\&$ Fabian~(\cite{All97}) found in
a sample of X-ray luminous clusters indications for a decrease of
$f_{\rm gas}$ with increasing $T_{\rm gas}$. Ettori $\&$ Fabian~(\cite{Ett99})
found no dependence of $f_{\rm gas}$ on $M_{\rm tot}$ for high luminosity
clusters.

As mentioned before the gas mass fraction is not constant with
radius. In order to plot this increase against other cluster properties, we compare the gas mass fraction at radius $r_{500}$ with the gas mass fraction at $0.5\times r_{500}$ in each cluster. The mean gas mass
fraction at $0.5\times r_{500}$ is $0.138\pm0.016$, i.e. smaller than the mean of $0.16\pm0.02$ at $r_{500}$.
The ratio of these fractions is a measure for the extent of the gas distribution relative to the dark matter extent: $E=\frac{f_{\rm gas}(r_{500})}{f_{\rm gas}(r_{500}/2)}$.

The error in the relative gas extent $\it{E}$ is not easy to determine. We estimate its uncertainty by testing how much this value changes when:
\begin{itemize}
\renewcommand{\labelitemi}{$\bullet$}
\item{we consider the uncertainty in the gas temperature}
\item{the fit parameters change}
\end{itemize}

Since the relative gas extent is calculated as the ratio of gas mass
fractions at different radii, the errors coming
from the uncertainty in the temperature are cancelled out.
The relative gas extent $\it{E}$ changes significantly when the fit parameters ($\beta$ and $r_{\rm c}$) are varied. This error coming from the uncertainties in the fit parameters is included in the following graphs.

For all the clusters in our sample the relative gas extent
$E$ is larger
than 1 (see Fig.~\ref{fig:E_Mt}).
This means that in general the gas
distribution is more extended than the dark matter, which is in agreement with other results for nearby cluster samples by David et al.~(\cite{Dav95}), Jones $\&$ Forman~(\cite{Jon99}), Ettori $\&$ Fabian~(\cite{Ett99}) and for distant samples by Schindler~(\cite{Sch99}) and Tsuru et al.~(\cite{Tsu97}). The distant
sample by Schindler~(\cite{Sch99}) and the other Chandra and XMM observations are also shown in Fig.~\ref{fig:E_Mt} for comparison.

In the nearby sample ($4.7<T<9.4$ keV) this relative gas extent $\it{E}$ shows a mild
dependence on the total cluster mass (see Fig.~\ref{fig:E_Mt})
similar to the result by Schindler~(\cite{Sch99}). Clusters with larger masses tend to have smaller relative gas extents (similar dependence confirmed by Reiprich $\&$ B\"ohringer~\cite{Rei99} although they used different radii). 

\begin{figure}[ht]
\centering
\includegraphics[width=10cm]{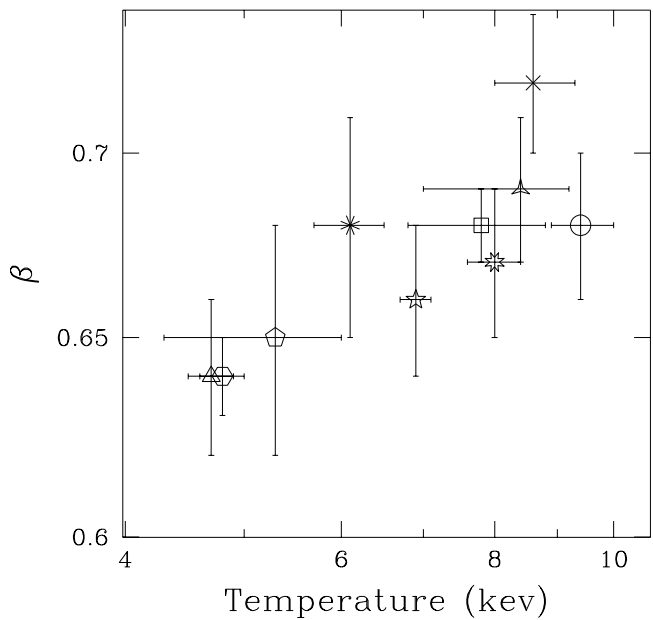}
\caption{$\beta$-temperature relation. $\beta$ is the parameter obtained when fitting with a $\beta-$model the X-ray surface brightness profile. There is a trend to find larger $\beta$ values for larger temperatures.}
\label{fig:beta_T}
\end{figure}
The differences in the gas and dark matter distribution cannot be explained by purely gravitational energy coming from the collapse of the cluster. Additional energy input is necessary to explain it, e.g. from supernova-driven galactic winds. It might be that this additional energy affects low mass clusters more than massive clusters, so that a massive cluster can maintain a ratio $E\approx1$, while in the smaller clusters the gas is more extended. 

Entropy studies by Ponman et al.~(\cite{Pon99}) of cool clusters ($T<4$keV) observed with ROSAT and GINGA suggested that for these systems, (pre-)heating processes play an important role in cluster formation.

Equation~(\ref{eq:mass}) shows that for a given radius $M_{\rm tot}\propto T \times \beta$. If there is a temperature rise due to preheating processes, a cluster with a certain $M_{\rm tot}$ should have shallower gas density profile or more extended gas distribution ($\beta$ value smaller). Then if that heating is more effective in cooler cluster, i.e. less massive clusters, one should expect an anticorrelation between total mass and gas extent and lower values of $\beta$ in cooler systems. This seems to be consistent with observations (e.g. Mohr $\&$ Evrard~\cite{Mohr97}, Vikhlinin et al.~\cite{Vik99}, Ponman et al.~\cite{Pon99}, Nevalainen et al.~\cite{Neva00}). Our sample also exhibits this behavior, see Fig.~\ref{fig:beta_T}.

\subsection{Mass-temperature relation}
Accurate measurements of the cluster total mass are possible only for a
limited number of clusters. For this reason there are currently
not enough data available for a direct derivation of the mass function, which is crucial for the determinations of the cosmological
parameters using the cluster abundances at different redshifts. A more
practical way of determining the mass function is to observe the
distribution of readily available average cluster gas temperatures and
to convert these to masses, taking advantage of the tight
correlation between mass and temperature predicted by hydrodynamic cluster
formation simulations (e.g. Evrard et al.~\cite{Ev96}). Therefore,
a well-established $M_{\rm tot}-T$ relation can be used as a powerful
cluster mass estimator. Furthermore the $M_{\rm tot}-T$ relation is also
interesting in itself, because deviations from the predicted
self-similar scaling of $M\propto T^{3/2}$ would indicate that
more physical processes are at play than gravity alone.

\begin{figure*}[ht]
\begin{center}
\includegraphics[width=17cm]{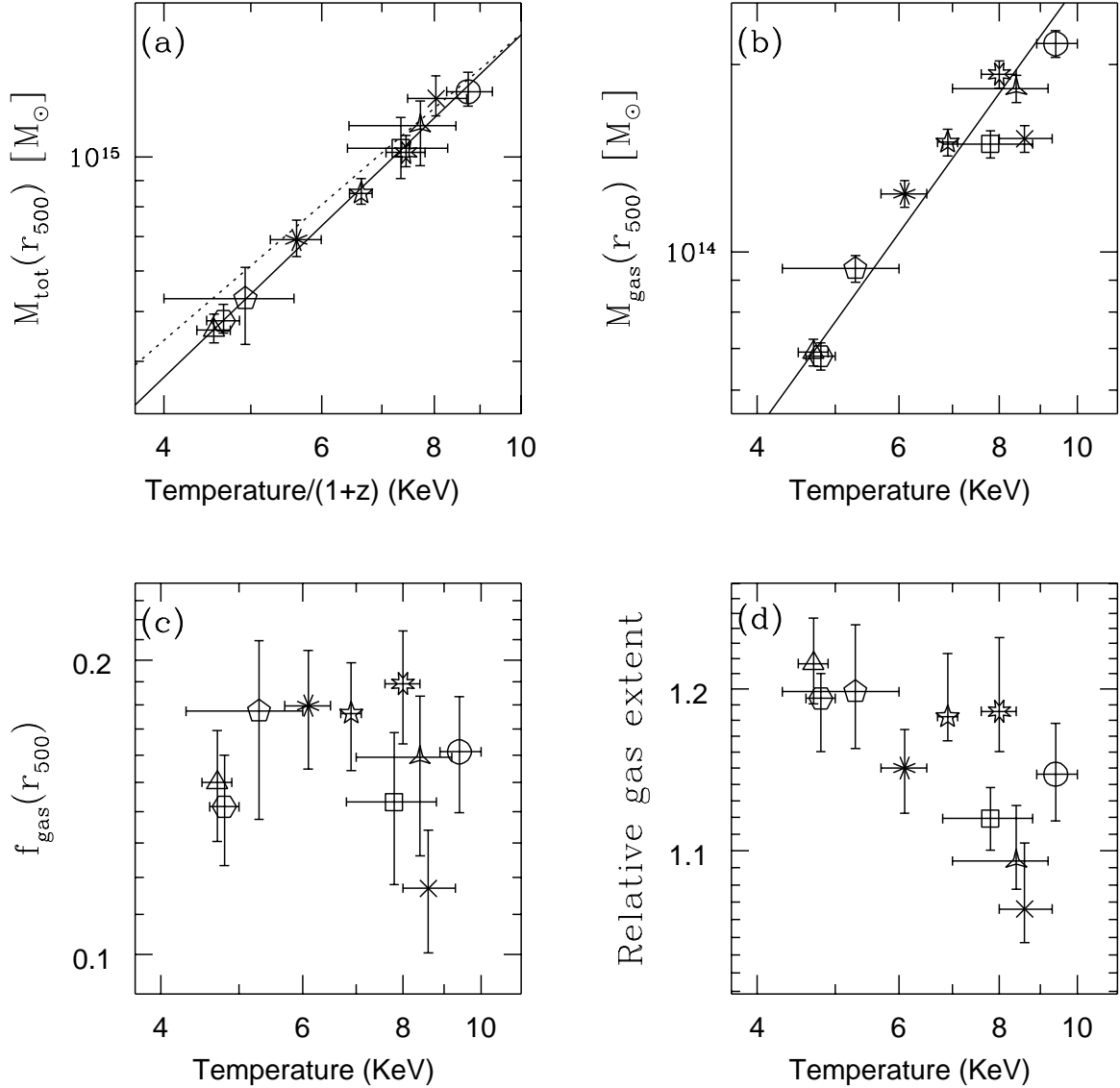}
\caption {Various quantities versus temperature. (a) Total mass. (b) Gas mass. Solid 
lines show the best fit for our sample. The dashed line is not a fit but the curve
expected from virial considerations with an arbitrary
normalisation. (c) Gas mass fraction. (d) Ratio of gas mass fractions at $r_{500}$ and $r_{500}/2$.}
\label{fig:relT}
\end{center}
\end{figure*}

Assuming self-similarity and a velocity dispersion proportional to the 
X-ray temperature, the virial theorem provides a relation between
total mass, radius and X-ray temperature:
$M_{\rm tot}(r_{500})/r_{500}\propto T$. Equivalently, $r_{500}$ can be expressed by the definition of the overdensity~$r_{500}\propto~M_{\rm tot}(r_{500})^{1/3}/(1+z)$ yielding the
relation $M_{\rm tot}(r_{500})~\propto (T/(1+z))^{3/2}$. We see this correlation
in our data (see Fig.~\ref{fig:relT}a). A fit taking into account the errors in the temperature and the error in 
the mass yields the following relation
$M_{\rm tot}(r_{500})=0.36{\left(\frac{T}{1+z}\right)}^{(1.7\pm0.2)}$ shown in Fig.~\ref{fig:relT}a as solid line. $M_{\rm tot}(r_{500})$ is in units of
$10^{14}M_{\odot}$ and T in units of keV. The slope is greater than
the virial value of 3/2 but consistent within the errors. If we do the same fit using the masses calculated with the temperature
profiles the slope found is $1.4\pm0.2$.

Other observational analyses, with only high-temperature ($kT>4-5$keV) clusters, have achieved results consistent with the
theoretical prediction (Hjorth et al.~\cite{Hjo98}, Neumann $\&$ Arnaud~\cite{Neu99}).

However many studies (e.g. Ponman et al.~\cite{Pon99}, Horner
et al.~\cite{Hor99}, Nevalainen et al.~\cite{Neva00}, Finoguenov et al.~\cite{Fin01}) have shown that the influence
of energy feed-back into ICM in low-temperature clusters/groups
can become more significant than that in high-temperature
systems. Consequently, the self-similarity may break at the
low-temperature end. If supernovae release a similar amount of energy
per unit gas mass in hot and cool clusters, the coolest clusters would
be affected more significantly and exhibit a stronger shift to higher
temperatures in the M-T diagram than the hotter
clusters. This will steepen the M-T relation.

Furthermore, we test the relation between the gas mass and 
gas mass fraction and the temperature. As expected from the gas mass$-$total mass relation, there is also a relation between gas mass and
temperature (see Fig.~\ref{fig:relT}b). A linear regression fit
yields $M_{\rm gas}(r_{500})=0.04T^{(1.80\pm0.16)}$ with $M_{\rm gas}(r_{500})$ in units
of $10^{14}M_{\odot}$ and T in keV. This correlation is in
agreement with other results in nearby clusters (Reiprich $\&$ B\"{o}hringer~\cite{Rei99};
Jones $\&$ Forman~\cite{Jon99}). For comparison, Reiprich $\&$ B\"{o}hringer~(\cite{Rei99}) find an
exponent of 2.08 and Schindler~(\cite{Sch99}) a larger exponent
4.1$\pm1.5$ for distant clusters. 

As expected from the non-correlation of the gas mass fraction with the
total mass, we find also no correlation between the gas mass fraction
and the temperature (see Fig.~\ref{fig:relT}c). This result is in good
agreement with Mohr et al.~(\cite{Mohr99}). They find a mild dependence
comparing low temperature clusters ($T<5$keV) with high temperature
clusters ($T>5$keV). For the high temperature clusters alone, in which 
category all our clusters fall, they find no dependence.

We find an interesting correlation between the relative gas extent and 
the temperature (see Fig.~\ref{fig:relT}d), which is of course related to 
the dependence of the relative gas extent on the total mass, shown
above. The relative gas extent tends to be slightly larger in lower temperature clusters.


\section{Summary}

We have analysed a sample of ten nearby clusters of galaxies using
the X-ray data provided by the ROSAT and ASCA satellites. For this sample
physical quantities like gas mass, total mass, gas mass fraction and
relative gas extent have been derived. Correlations between the above quantities have
been studied and our findings are
\begin{itemize}
\renewcommand{\labelitemi}{$\bullet$}
\item{The gas mass fraction increases with radius for all our clusters
implying that gas is more extended than dark matter,  confirming previous results (David et al.~\cite{Dav95}, Ettori \& Fabian~\cite{Ett99}, Jones $\&$ Forman~\cite{Jon99}). This behaviour is
more pronounced when temperature profiles are taking into account in the mass
analysis.}
\item{Within $r_{500}$, the mean gas mass fraction obtained is $(0.16\pm0.02)h_{50}^{-3/2}$.

We see no trend in the gas mass fraction with redshift.}
\item{The gas extent relative to the dark matter distribution shows a mild dependence on the total
mass and gas temperature. Clusters with larger masses have smaller
relative gas extents, as we would expect if non-gravitational processes
are important in cluster formation. Hints for this kind of behaviour have been found previously in distant clusters. These new results confirm this trend.}

\item{Studying the mass-temperature relation we find a slope slightly
steeper compared with the theoretical value of 3/2 although consistent
within the errors. The self-similar slope 3/2 is found when using temperature
gradient analysis. Other observational analyses, with only high-temperature clusters, have achieved results consistent with the theoretical prediction (Hjorth et al.~\cite{Hjo98}, Neumann $\&$ Arnaud~\cite{Neu99}, Nevalainen et al.~\cite{Neva00}).}

\end{itemize}


\begin{acknowledgements} 

A. Castillo-Morales acknowledges support by the Marie Curie Training Site grant HPMT-CT-2000-00136
by the European Comission.
\end{acknowledgements}

\end{document}